\documentclass[twocolumn,superscriptaddress,aps,preprintnumbers,amsmath,amssymb,prl,nofootinbib]{revtex4-1}
\usepackage{amsmath,graphics,epsfig}
\usepackage{color,hyperref}
\usepackage{datetime}
\begin{document}

\title{Natural Supersymmetry From Dynamically Reduced Radiative Correction}
\author{Huayong Han}
\affiliation{Department of Physics, Chongqing University, Chongqing 401331, P.R. China}
\author{ Sibo Zheng}
\affiliation{Department of Physics, Chongqing University, Chongqing 401331, P.R. China}

\date{June, 2015}

\begin{abstract}
New natural supersymmetry is explored in the light of dynamically reduced radiative correction.
Unlike in the conventional natural supersymmetry,
the range of supersymmetric mass spectrum can be far above the TeV scale instead.
For the illustrating model of non-universal gaugino masses,
the parameter space which satisfies the Higgs mass and other LHC constraints
is shown explicitly.
We propose that this example can be realized by employing the no-scale supergravity.
\end{abstract}

\maketitle

\section{1.~Introduction}
Following our previous works \cite{Zheng:2015, Zheng:2013},
we continue to explore new natural supersymmetry (SUSY)
beyond the conventional situation,
the latter of which is severely challenged by the present LHC data.
The conventional natural SUSY is heavily based on the understanding of the naturalness,
which means that the weak scale should be not sensitive to the variation of the fundamental SUSY parameters labelled by $a_i$.
The sensitivity can be measured by defining a fine tuning parameter $c= {\rm max}\{\mid c_{i}\mid\}$
\cite{Ellis:1986yg, Barbieri:1987fn},
where
\begin{eqnarray}
c_{i}\equiv\partial {\rm ln~M_{Z}^2}/ \partial {\rm ln~a_{i}}\sim \partial {\rm ln~m_{H_{u}}^2}/ \partial {\rm ln~a_{i}}.\nonumber
\end{eqnarray}
The requirement that the fine-tuning should not be too large,
for example $c \leq 100$,
leads to natural SUSY mass spectrum beneath TeV scale.
Unfortunately, the present LHC data, especially about the Higgs mass \cite{125:ATLAS, 125:CMS}
and the lower bounds \cite{ATLAS2013,CMS2013} on SUSY particle masses,
disfavors such conventional natural SUSY in the context of the minimal supersymmetric standard model (MSSM).
This is the main motivation for the study of new natural SUSY.

Among the efforts to construct new natural SUSY,
re-understanding the upper bounds on SUSY mass spectrum due to the naturalness argument
is an interesting direction.
As SUSY mass spectrum is usually obtained
through messengers that transmit the SUSY breaking effects in the SUSY-breaking sector to the MSSM,
it actually depends on few mass parameters less than one expects naively,
namely the SUSY-breaking scale and the SUSY-breaking mediation scale.
In this sense, the upper bounds on natural SUSY mass spectrum are over estimated.
Along this line it was shown in \cite{9908309, 9909334} that
these upper bounds are relaxed to
multi-TeV scale in the light of cancellations among the large renormalization group (RG) effects.
For more recent discussions, see \cite{Zheng:2015} .

The observation above can be extended to broader phenomenological applications.
Our technical strategy, which is viable to any well defined and perturbatively valid SUSY mode,
is outlined as follows.\\
$(a)$, Parameterize the SUSY mass spectrum at the input scale with a few unfixed parameters,
which define the parameter space.\\
$(b)$, Generate the SUSY mass spectrum at the weak scale with SUSPECT \cite{Djouadi:2002ze}
or other similar codes.\\
$(c)$, Impose the condition arising from dynamically reduced radiative correction (DRRC),
\begin{equation}
\label{naturalness}
-(300~{\rm GeV})^2 \leq m_{H_u}^2[M_Z] \leq -(200~{\rm GeV})^2,
\end{equation}
which implies that cancellations among various large RG corrections to the up-type Higgs mass squared indeed happen.
The range in Eq.(\ref{naturalness}) can be slightly adjusted.\\
$(d)$, Impose the experimental constraints such as Higgs mass, dark matter (DM) relic density,
lower bounds on SUSY particle masses, etc.

We will use the scenario of gaugino masses
arising from the no-scale supergravity \cite{Ellis:1983, Ellis:1984}  as an illustrating example \footnote{For a review, see, e.g., \cite{Lahanas:1986uc}.}.
In this model
the mass parameters at the input scale include three gaugino masses $M_i$ and $\mu$ term \cite{Ellis:1985jn, Drees:1985bx},
all of which may only depend on the single SUSY-breaking scale
once the mediation scale is identified.
For example, if the input scale is chosen as the grand unification (GUT) scale
the number of mass parameters may reduce to one in simplified model,
in which one obtains universal gaugino masses $M_{i}=M_{\text{universal}}$.
However, this simplest case is excluded by the condition of DRRC in Eq.(\ref{naturalness}).

Our parameterization for the parameter space is similar to some of earlier literature \cite{Hotta:1995cd, ArkaniHamed:1996jq, Kurosawa:1999bh,Huitu:1999vx, Anderson:1999uia,Komine:2000tj, Abe:2007kf, 0908.0857}.
However, there are three important differences,
which are the new condition of DRRC in Eq.(\ref{naturalness}),
the latest LHC constraints in Eq.(\ref{constraint}), and a different choice on the input scale $M$.

The paper is organized as follows.
In section 2,  we scan the parameter space in the context of MSSM
where we use SUSPECT \cite{Djouadi:2002ze} to calculate the Higgs mass at the 2-loop level
\footnote{As shown in  \cite{Feng:2013tvd}
there is about $\sim 0.5-2$ GeV difference between the 2-loop and 3-loop result when the SUSY mass spectrum is above 3 TeV in the light of $\text{H3m}$,
which is a fortran code of 3-loop Higgs mass calculation.
The parameter space is not dramatically affected by the 2-loop approximation.},
and MicrOMEGAs 4.1.8 \cite{Belanger:2014vza} to calculate the relic density of light and neutral higgsino \footnote{
For wino and bino masses far above 1 TeV but small $\mu$ term,
the neutral higgsino state mainly serves as the DM candidate.}.
In section 3, we discuss the model building for the parameter space shown in the section 2 by employing the no-scale supergravity.
Finally, we conclude in section 4.

\section{2.~Constraints and Parameter Space}
This section to devoted to explore the parameter space for the model of non-universal gaugino masses, which is parameterized by the three gaugino masses, $\mu$ term and $\tan\beta$,
\begin{eqnarray}
\label{parameterspace}
\{M_{1},~M_{2},~M_{3},~\mu,~\tan\beta\}.
\end{eqnarray}
The following constraints are imposed in the light of the LHC data about the Higgs mass ~\cite{125:ATLAS, 125:CMS} and SUSY mass bounds~\cite{ATLAS2013, CMS2013},
\begin{eqnarray}
\label{constraint}
\text{Higgs mass}: 125.3< m_{h} < 126.1~\text{GeV}, \nonumber\\
\text{gluino mass bound}: m_{\tilde g} > 1.3~\text{GeV}, \nonumber\\
\text{light stop mass bound}:m_{\tilde t_1} > 600~\text{GeV}, \nonumber\\
\text{Other scalar mass bounds}: m_{\tilde f} > 1~\text{TeV}.
\end{eqnarray}
The last constraint in Eq.(\ref{constraint}) is expected to be one of main results in DRRC induced natural SUSY, which obviously differs from the conventional natural SUSY mass spectrum.
The constraints in Eq.(\ref{constraint}) guarantee all present LHC data about SUSY can be easily explained.

Once the input mass scale $M$ is chosen,
the SUSY mass spectrum at the weak scale can be calculated.
Instead of the choice on $M\sim 10^{16}$ GeV in some of previous works,
we take $M$ as the Plank mass scale $M_{\text{P}}=2.4\times10^{18} \, {\rm GeV}$.
The main reason for this different choice is that
we would like to obtain large mass splitting among the gaugino masses other than the universal case.
This can be realized in various ways.
For instance, when the higher order corrections \cite{Drees:1985bx} to the gaugino masses are not small, which implies that in the context of no-scale supergravity $M$ must be close to the Plank mass,
one may obtain large mass splittings among gaugino masses
\footnote{We refer the reader to section 3 for more discussions about the ways in which one may obtain large splitting among gaugino masses.}.

In terms of the code S{\small U}S{\small PECT} 2.41 \cite{Djouadi:2002ze}
we show in Fig.\ref{fig1} the parameter space which satisfies DRRC in Eq.(\ref{naturalness})
and all the constraints in Eq.(\ref{constraint}) simultaneously in the range,
\begin{eqnarray}
0< M_{3} &\leq& M_{1},\nonumber\\
0< M_{2} &\leq& M_{1}, \nonumber\\
5~\text{TeV} < M_{1} &<& 30~\text{TeV}.
\end{eqnarray}
We have used two dimensional real numbers $x$ and $y$ to parameterize the gaugino masses,
\begin{eqnarray}
\label{xy}
\{M_{1},~M_{2},~M_{3}\}=M_{1}\{1,~x,~y\}.
\end{eqnarray}
The pattern in Fig.\ref{fig1} indicates that similar to the case of focus point \cite{9908309, 9909334, Zheng:2015}
cancellations among large RG corrections to $m^{2}_{H_{u}}$ indeed happens.
In order to satisfy the condition of DRRC, the bino mass is always the biggest one among the three gaugino masses,
and the ratios $M_{2}/M_{1}$ and $M_{3}/M_{1}$ both decrease as $M_{1}$ increases.

As $M_{i}$ are far above the weak scale,
the neutral higgsino can serve as the DM candidate,
which should not excess the relic density measured by the combination of Plank and WIMP 9-year data~\cite{Ade:2013zuv},
\begin{eqnarray}
\Omega \text{h}^{2}_{\text{higgsino}}\leq 0.1199 \pm 0.0027.
\end{eqnarray}
We show in Fig.\ref{fig2} the relic density of light and neutral higgsino in terms of
MicrOMEGAs 4.1.8 \cite{Belanger:2014vza},
which indicates that it cannot fully accommodate DM relic density in the most of the parameter space.
This is consistent with an analytic approximation \cite{0908.0857} for the case of higgsino DM,
\begin{eqnarray}
\Omega \text{h}^{2}_{\text{higgsino}}\sim 0.1 \times \left(\frac{\mu}{1\text{TeV}}\right)^{2}.\nonumber
\end{eqnarray}

\begin{widetext}
\begin{center}
\begin{figure}
\centering
\begin{minipage}[b]{0.5\textwidth}
\centering
\includegraphics[width=3in]{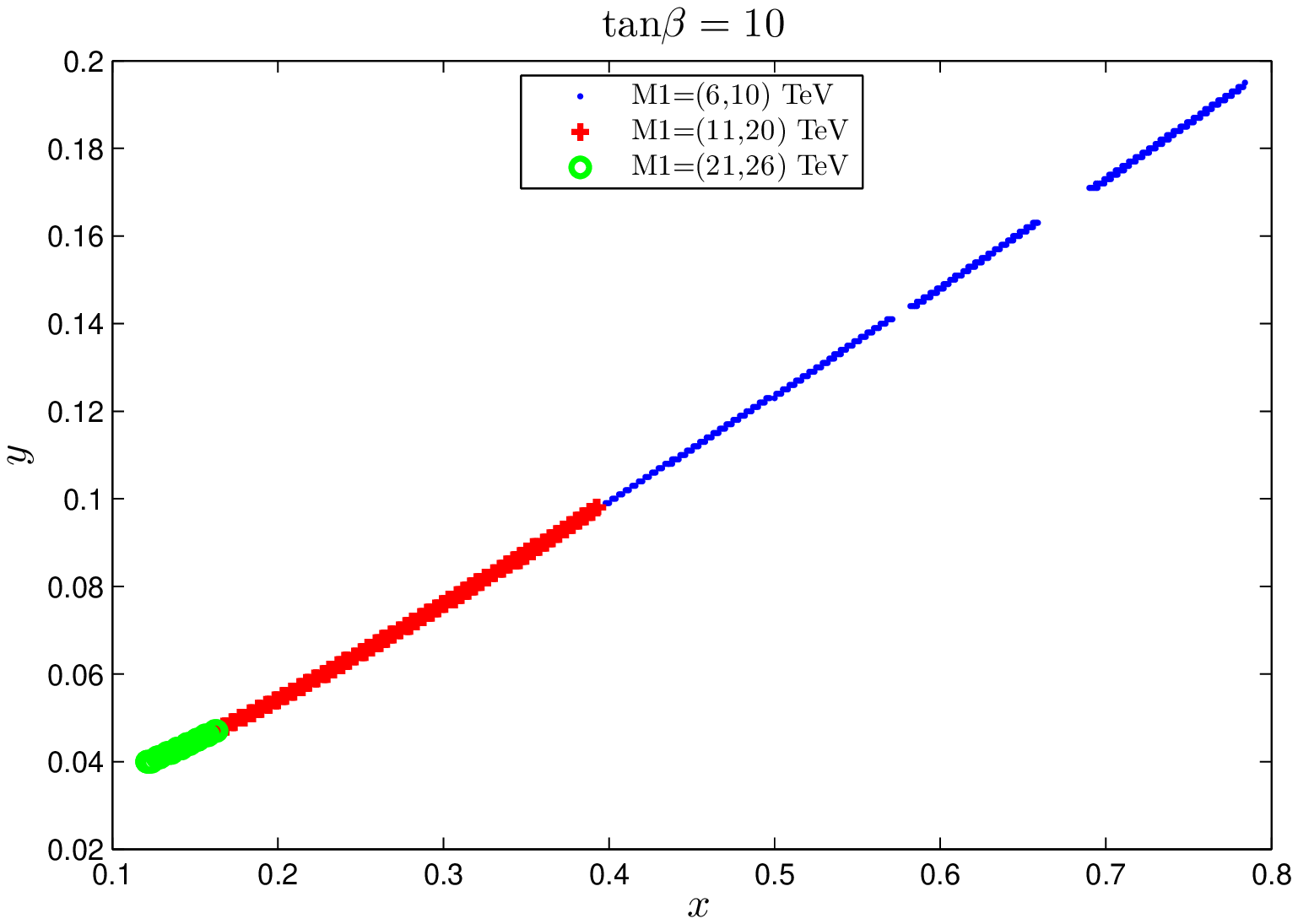}
\end{minipage}%
\centering
\begin{minipage}[b]{0.5\textwidth}
\centering
\includegraphics[width=3in]{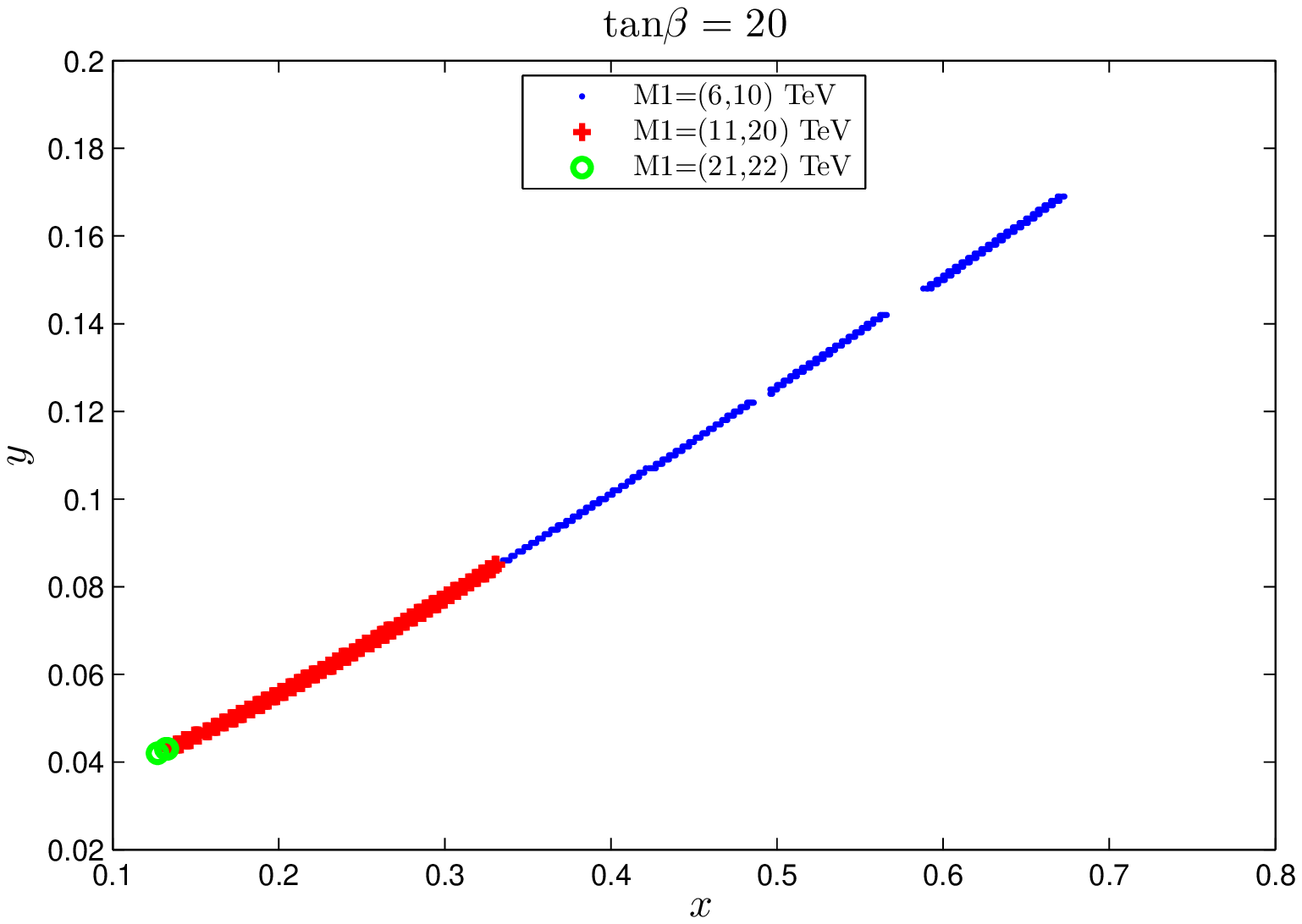}
\end{minipage}%
\caption{Parameter space projected to the plane of $x$ and $y$ for different ranges for $M_{1}$, which satisfies all the constraints in Eq.(\ref{constraint}) and DRRC in Eq.(\ref{naturalness}) simultaneously. $\tan\beta=10$ (left) and 20 (right), respectively.}
\label{fig1}
\end{figure}
\end{center}
\end{widetext}

As shown in the right plot in Fig.\ref{fig2},
this estimate changes mildly for a different value of $\tan\beta$,
thus more sensitive to the value of $\mu$.
If one adopts more negative $m^{2}_{H_{u}}$ than in Eq.(\ref{naturalness}),
larger $\mu$ will be induced in terms of  the conditions of electroweak symmetry breaking.
So, higgsino can accommodate DM relic density with the price of larger fine tuning.
Unlike in the context of MSSM,
this issue will be further discussed in the context of next-to-minimal supersymmetric model \cite{Han:2015}.

\begin{widetext}
\begin{center}
\begin{figure}
\centering
\begin{minipage}[b]{0.5\textwidth}
\centering
\includegraphics[width=3in]{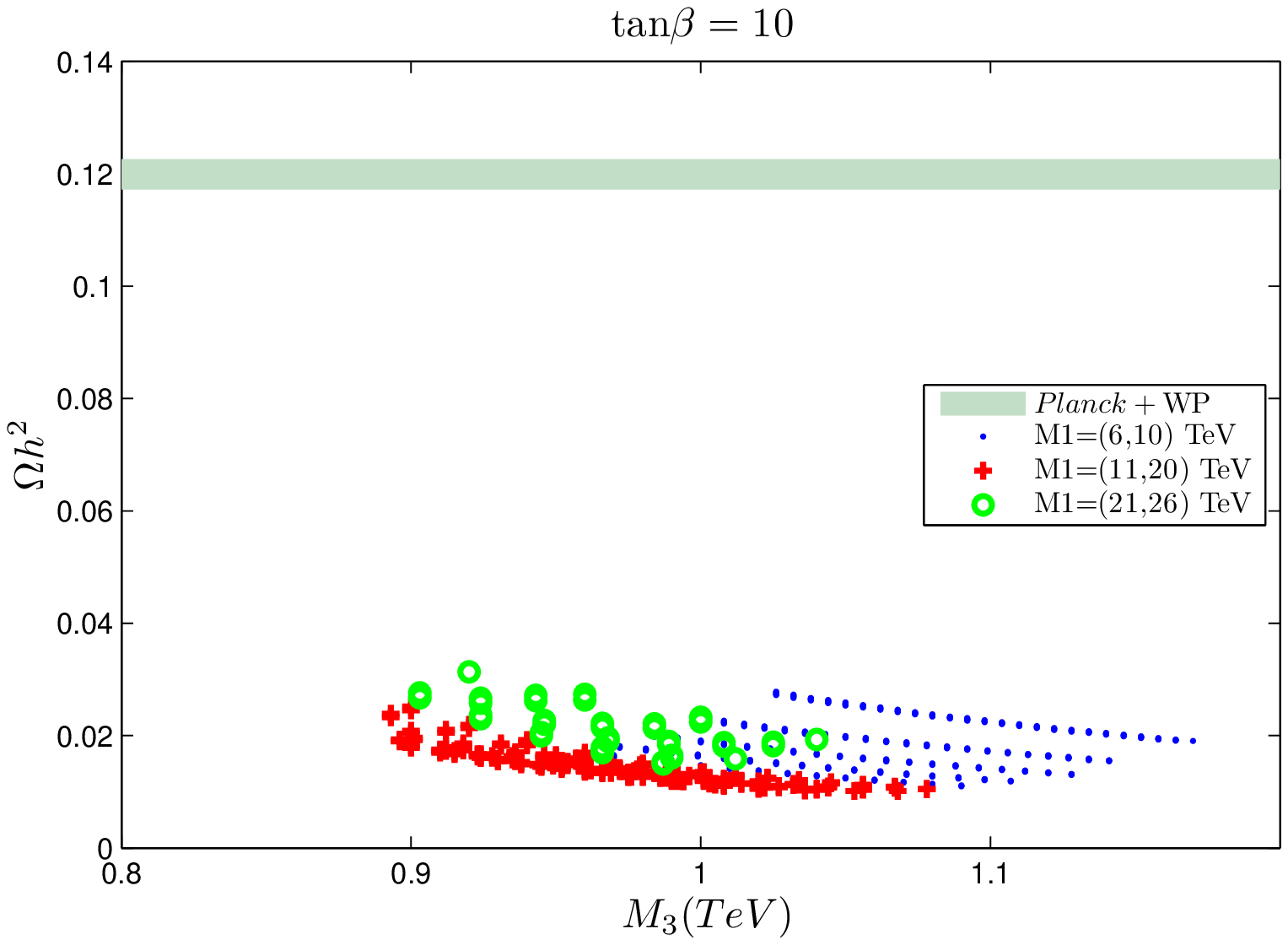}
\end{minipage}%
\centering
\begin{minipage}[b]{0.5\textwidth}
\centering
\includegraphics[width=3in]{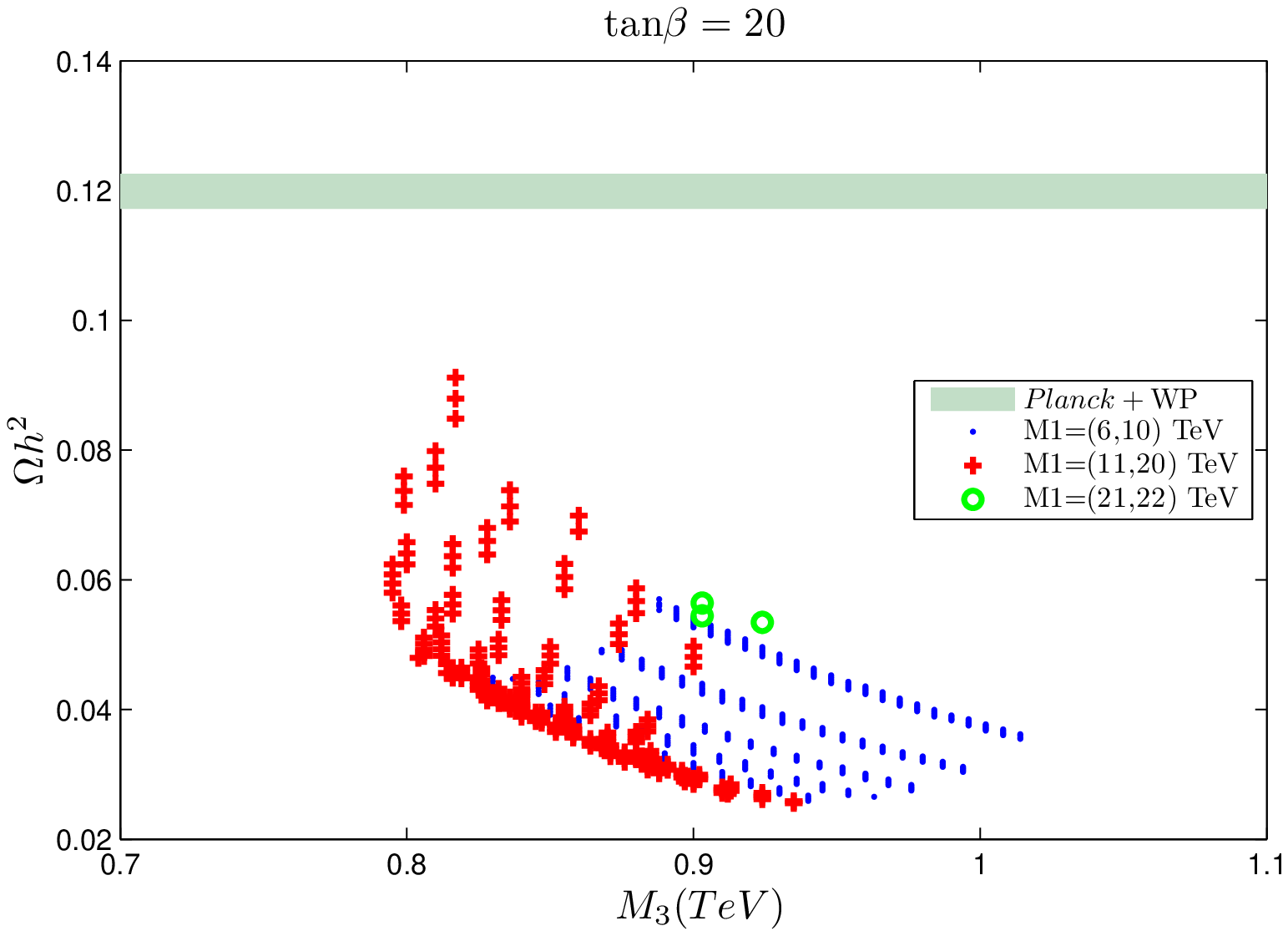}
\end{minipage}%
\caption{Neutral higgsino relic density as function of gluino mass $m_{\tilde{g}}[M]$ subtracted from Fig.\ref{fig2}. The green band corresponds to the value of DM relic density measured by the Plank and WIMP data \cite{Ade:2013zuv}. }
\label{fig2}
\end{figure}
\end{center}
\end{widetext}

\section{3.~Model Building}

In this section we employ the no-scale supergravity to construct the model of non-universal gaugino masses
with parameter space shown in Fig.\ref{fig1} and Fig.\ref{fig2}.
It was firstly shown in \cite{Ellis:1983, Ellis:1984}
that this theory can address the cosmological constant and mass hierarchy problems simultaneously.

We briefly review the phenomenology of no-scale supergravity.
We take the adjoint field under the representation $\mathbf{24}$ of $SU(5)$ for example,
the vacuum expectation value (vev) of which spontaneously breaks this $SU(5)$
into the SM gauge group $G_{\text{SM}}=SU(3)_{c}\times SU(2)_{L}\times U(1)_{Y}$.
The soft SUSY mass spectrum includes gaugino masses, which are given by,
\begin{eqnarray}
\label{gauginos}
\frac{M_1}{\alpha_1}&=&\frac{m_{3/2}}{\alpha_{5}\Gamma}\left(1-\xi_{1}+\xi_{2}\right), \nonumber \\
\frac{M_2}{\alpha_2}&=&\frac{m_{3/2}}{\alpha_{5}\Gamma}\left(1-3\xi_{1}\right), \nonumber\\
\frac{M_3}{\alpha_3}&=&\frac{m_{3/2}}{\alpha_{5}\Gamma} \left(1+2\xi_{1}\right),
\end{eqnarray}
where $m_{3/2}$ refers to the gravitino mass,
$\alpha_{i}$ $(i=1,2, 3)$ and $\alpha_{5}$
are the three fine structure constants of SM gauge coupling
and the fine structure constant of $SU(5)$ gauge group, respectively.
$\Gamma$, $\xi_{1}$ and $\xi_{2}$  are a set of functionals of
the vevs of the adjoint field(s) and scalars in the SUSY-breaking sector,
the definitions of which can be found in \cite{Ellis:1985jn}.
In other words, the information about the SUSY-breaking sector are restored in these three dimensionless quantities.

In particular,
the magnitude of $\xi_{1}$ and $\xi_{2}$ are controlled by the input mass scale $M$ relative to $M_{P}$.
If $M$ is far smaller than $M_{P}$ such as in GUT-scale SUSY models and $\xi_{1}$ is small,
they lead to an approximate relation from Eq.(\ref{gauginos}),
\begin{eqnarray}
\frac{M_3}{\alpha_{3}}\simeq\frac{M_2}{\alpha_{2}}\simeq\frac{M_1}{\alpha_{1}}.
\end{eqnarray}
Then the differences among $M_{i}$ at scale $M$ depend on the splitting of magnitudes
among $\alpha_{i}$ at the same scale.
This splittings are determined by the coefficients $f_i$ which are defined as,
\begin{equation}
\alpha_3[M]f_3 = \alpha_2 [M]f_2 =\alpha_1[M]f_1=\alpha_{5},
\end{equation}
The differences among $f_{i}$ are constrained by the measured values of gauge couplings
and the weak mixing angle $\sin^{2}\theta_{W}$ at the weak scale \cite{Drees:1985bx} .
One finds that the universal relation $f_{1}\simeq f_{2}\simeq f_{3}$ is approximately valid
for $M$ far below the Plank mass scale.
This implies that universal gaugino masses are expected.

The universal gaugino masses can be violated either
when $\xi_{1}$ is significant \cite{Ellis:1985jn},
the spontaneous breaking patterns of $SU(5)$ are other than $\mathbf{24}$ \cite{Ellis:1985jn, Huitu:1999vx}, the $SU(5)$ is extended into larger group structure \cite{Hotta:1995cd, ArkaniHamed:1996jq, Kurosawa:1999bh} or $M$ is close to $M_{P}$ \cite{Drees:1985bx}.
The last situation implies that $\xi_{2}$ which captures the leading order correction also becomes significant.
As chosen for the analysis in the previous section, we have in terms of Eq.(\ref{xy}) and Eq.(\ref{gauginos}) for the case of $M_{1}=20$ TeV,
\begin{eqnarray}
\label{xi}
\frac{1-3\xi_{1}}{1-\xi_{1}+\xi_{2}}&=&\frac{\alpha_{1}}{\alpha_{2}}x\simeq 0.12, \nonumber\\
\frac{1+2\xi_{1}}{1-\xi_{1}+\xi_{2}}&=&\frac{\alpha_{1}}{\alpha_{3}}y \simeq 0.04.
\end{eqnarray}
which implies that $\xi_{1}\simeq -0.22 $ and $\xi_{2}\simeq 13.9$.

Alternatively,  one may extend the $SU(5)$ into a larger group structure,
\begin{eqnarray}
 SU(5)\times G_{H},
\end{eqnarray}
where $G_{H}$ is the gauge group of hidden sector.
The spontaneous breaking scale of $G_{H}$ is assumed to be larger than $M$.
Given the general choice $G_{H}=SU(3)_{H}\times SU(2)_{H} \times U(1)_{H}$,
the SM gauge group $SU(3)_{c}$ is regarded as the diagonal subgroup of $SU(3)\times SU(3)_{H}$, etc,
where $H$ labels the hidden sector.
The gaugino masses are read from the Lagrangian in the gauge eigenstate formalism,
\begin{eqnarray}
\mathcal{L}_{\text{soft}}&\sim& m_{5}\lambda_{i}\lambda_{i}
+m_{H_{i}}\lambda_{H_{i}}\lambda_{H_{i}}.
\end{eqnarray}
which gives rise to,
\begin{eqnarray}
\frac{M_{1}}{\alpha_{1}}&=&\frac{m_{H_{1}}}{\alpha_{H_{1}}}+\frac{m_{5}}{\alpha_{5}},\nonumber\\
\frac{M_{2}}{\alpha_{2}}&=&\frac{m_{H_{2}}}{\alpha_{H_{2}}}+\frac{m_{5}}{\alpha_{5}},\nonumber\\
\frac{M_{3}}{\alpha_{3}}&=&\frac{m_{H_{3}}}{\alpha_{H_{3}}}+\frac{m_{5}}{\alpha_{5}}.
\end{eqnarray}
Assume the magnitudes of all gauge couplings of same order $\alpha_{H_{i}}\sim \alpha_{5}$.
The large mass splittings as required in the previous section can be explained by
\begin{eqnarray}
m_{5} \sim m_{H_{3}}\sim \frac{1}{15}m_{H_{1}}\sim \frac{1}{3} m_{H_{2}}
\end{eqnarray}

\section{4.~Conclusion}
In this paper,  the scheme for a new type of natural SUSY is proposed.
This scheme is valid for models in which the SUSY mass spectrum at the input scale depdends on
a single fundamental mass scale.
It is usually true when the mediation scale of SUSY breaking
is identified as the spontaneous breaking scale of some hidden gauge symmetry.

As an illustration for the scheme,
we employ the scenario of non-universal gaugino masses,
which can be easily constructed in the context of no-scale supergravity.
No-scale supergravity is a typical example
in which the mediation scale of SUSY breaking is set by the spontaneous breaking scale
of the $SU(5)$ gauge symmetry.
We have shown that in such model
SUSY mass spectrum far above the TeV scale is indeed natural,
and that a small part of parameter space can explain the Higgs mass constraint
and all present LHC data about SUSY.\\

~~~~~~~~~~~~~~~~~~~
$\mathbf{Acknowledgement}$\\
This work is supported in part by National Natural Science Foundation of China under Grant No.11247031, 11405015, and 11447199.

\end{document}